# Strengthening of foamed composite materials

Y. Khidas[1], B. Haffner[2] and O. Pitois[2]

[1] *Université Paris Est, Laboratoire Navier, UMR 8205 CNRS – École des Ponts ParisTech – IFSTTAR, 5 bd Descartes, 77454 Marne-la-Vallée Cedex 2, France*

[2] *Université Paris Est, Laboratoire Navier, UMR 8205 CNRS – École des Ponts ParisTech – IFSTTAR, cité Descartes, 2 allée Kepler, 77420 Champs-sur-Marne, France*

**Abstract:**

We investigate the shear elastic modulus of soft polymer foams loaded with hard spherical particles and we show that, for constant bubble size and gas volume fraction, strengthening is strongly dependent on the size of those inclusions. Through an accurate control of the ratio $\lambda$ that compares the particle size to the thickness of the struts in the foam structure, we evidence a transition in the mechanical behavior at $\lambda \approx 1$. For $\lambda < 1$, every particle loading leads to a strengthening effect whose magnitude depends only on the particle volume fraction. On the contrary, for $\lambda > 1$, the strengthening effect weakens abruptly as a function of $\lambda$ and a softening effect is even observed for $\lambda \gtrsim 10$. This transition in the mechanical behavior is reminiscent of the so-called "*particle exclusion transition*" that has been recently reported within the framework of drainage of foamy granular suspensions [Haffner B, Khidas Y, Pitois O. The drainage of foamy granular suspensions. J Colloid Interface Sci 2015. In Press.]. It involves the evolution for the geometrical configuration of the particles with respect to the foam network, and it appears to control the mechanics of such foamy systems.

1. **Introduction**

The incorporation of particles into materials is a well-known strategy to improve their mechanical behavior. When dealing with foamy materials, one has to consider the typical size of the struts that form the solid skeleton compared to the filler size. Several studies have been devoted to the reinforcement of such foams. For example, polyurethane foams were filled with glass fibres [1], metallic powders [2], organic particles [3] or mineral particles [4–7], … Aerated cementitious materials also belong to the class of foamed composite materials. In the current climate of sustainable development, those foam-based materials are destined to expand and in certain cases to replace advantageously conventional building materials. Numerous works were focused on those materials in order to improve their mechanics [8,9] by adding different types of filler, such as sand [10], fly ashes [10,11], glass fibers [12], …

One of the essential features of foamy composite materials is that the filler size has to be fine enough for strengthening to be observed [2–5,10], and it has been shown that reinforcement is not efficient if the added particles are bigger than the bubbles [7]. The effect of the filler size has been attributed to the fact that only the small particles can enter the struts that form the foam skeleton. Therefore, the distinction that can be made for the mechanics of foams involving "small" and "large" particles is necessary related to distinct geometrical configurations for the particles within the foam. Surprisingly, whereas this distinction is expected to be based on geometry only, it has never been studied from the purely geometrical point of view. Indeed, most of reported results correspond to the global effect of the filler, including filler-matrix interaction [4] and complex interaction in the chemical reaction from which bubbles are generated [4,6,13]. Moreover, the bubble size and the gas volume fraction cannot be kept constant when studying the effect of the loading, which increases the difficulty to disentangle all types of effect. Consequently, our understanding of the geometrical effect for filler particles in solid foams is rather qualitative, whereas the engineering of such complex systems requires a quantitative estimate for this effect. Clearly, a suitable approach would consist in using perfectly inert particles and changing the particle loading only, all relevant parameters of the reference foam, i.e. the unloaded foam, being unchanged.

In this paper, we propose a generic approach to better understand the subtle interplay between the complex geometry of particulate foams and their mechanics. We investigate the shear elastic modulus of systems which consist of soft polymer foams loaded with hard spherical particles. The experimental study is conducted in such a way that monodisperse precursor foams are unaltered by the loading process, allowing us to assess the strengthening effect due solely to the presence of the particles.

## 2. Experimental

2.1 sample preparation

Our model system is a soft polymer foam loaded with hard spherical inclusions. The soft polymer is a porcine gelatin. The gelation is thermoreversible: above $T_{gel} \approx$ 29°C gelatin is a liquid solution and below it becomes a soft elastic solid. The gelatin powder was kindly supplied by Rousselot and used as received. The inclusions are polystyrene beads (dynoseeds® from microbeads SA) characterized by a week polydispersity (standard deviation of particle size distributions around 5%). The large contrast in the elastic moduli of gelatin (few kPa) and polystyrene (few Gpa) will allow us to consider the beads as hard spheres.

In addition to gelatin, the foaming solution is composed of distilled water, glycerol to match the density of the beads (1.05) and trimethyl(tetradecyl)azanium bromide (TTAB) as a surfactant. Glycerol and TTAB were purchased from Aldrich and used as received. For the preparation, an aqueous solution of TTAB and glycerol is first prepared. Then, appropriated amounts of gelatin powder were dissolved under continuous stirring in the previous TTAB aqueous solution at 65 °C during 30 min to reach the final concentrations in gelatin 13.5 wt%, TTAB 0.3 wt% (well above the CMC of 1 g/L) and glycerol 12 wt%. After ultrasonic degassing during 1 min, the solution is kept at 50°C during 15 h. In the foaming solution, the polystyrene particles behave as fully hydrophilic particles and they do not adsorb at bubble interfaces.

The Fig. 1 is an overview of the experimental procedure used to generate the particle laden foams. Inside a box maintained at 60°C, the liquid gelatin solution is mixed with nitrogen in a T-junction to produce a precursor monodisperse foam. The gas fraction of this foam is set by controlling the flow rates of the liquid and the gas, using a syringe pump and a mass flow controller respectively. In a second step, a suspension of particles in liquid gelatin is pushed and mixed to the precursor foam thanks to a T-junction. The resulting particulate foam is pushed in an open cell composed of two transparent slabs attached to two spacers that impose the gap (Fig. 2). We rapidly quench the system in a refrigerator at 5°C during 5 min and then wait 55 min more at room temperature before the mechanical test. It is crucial that all the samples have the same temperature history. Indeed gelatin gels are known to exhibit logarithmic structural aging that leads to a natural increase of their elastic moduli [14]. In the same time, some water evaporation also contribute to make the system evolve, so that it is essential to follow the same protocol in order to have the required reproducibility on the mechanical results.

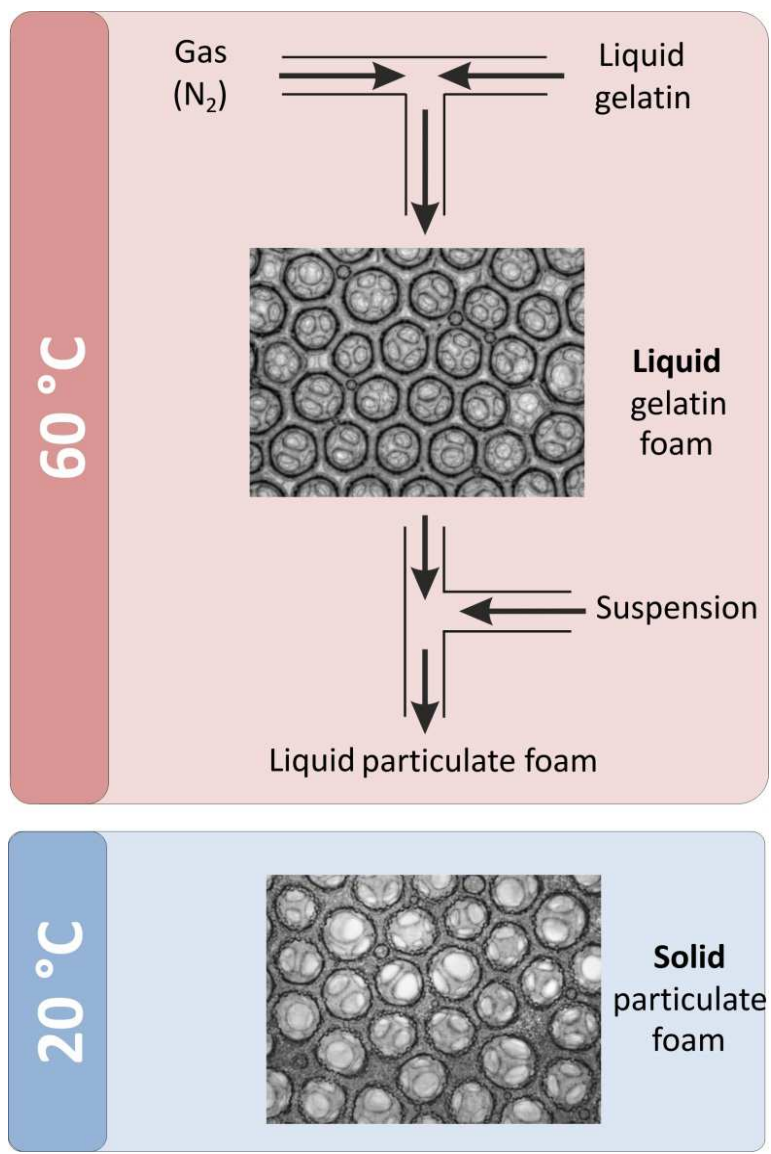

Fig. 1. Sketch of the experimental procedure to generate solid particulate foams with monodisperse bubbles and monodisperse particles.

In the present study, the gas volume fraction of the particle-laden foam is kept constant: $\phi = 0.8$. The particle size and the particle volume fraction are varied within the respective ranges: $d_p = 6 - 500$ µm and $\phi_p = 0 - 6$ %. $\phi_p$ corresponds to the particle volume over whole volume of foam, but it will be useful to define the particle volume fraction with respect to the foam skeleton: $\varphi_p = \phi_p/(1-\phi) = 0 - 30$ %. In order to study the effect due solely to the presence of particles in the foam, it is of first importance that making $d_p$ and $\phi_p$ vary do not modify the bubble size. Thanks to a microscope and to image analysis, we have checked that the bubble monodispersity is not degraded by the loading, and the bubble diameter was measured to be constant and equal to $D_b = 400 \pm 20$ µm whatever the loading level. The use of fluorescent particles enables us to check that particles are uniformly distributed in the whole foam. Note that particles are always larger than the thickness of the foam films, so that they are excluded from the latter rapidly after the generation stage. Particles were observed to be well-distributed in nodes and plateau. The typical size of our foam sample is 70 x 25 x 7.7 mm$^3$.

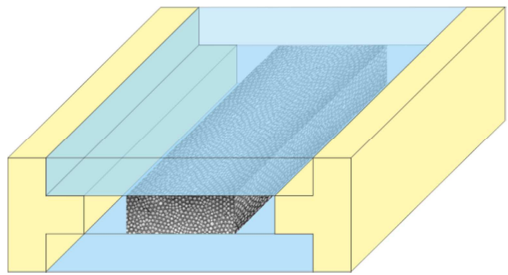

Fig. 2. The foam is poured in a cell composed of two transparent slabs with a controlled gap thanks to two spacers. For sake of clarity, the fixing device between the slabs and the spacers is not represented.

2.2 Plane shear test

Mechanical properties of our sample are characterized by the shear elastic modulus measured in a plane shear test. To shear the foam sample, one of the slabs of the cell is attached to a motorized translation stage (Newport ILS50 and its motion controller ESP301) while the other slab is maintained at a fixed position thanks to the below-balance weighing of a precision balance (Sartorius BP211D). This balance measures the vertical force required to maintain the slab motionless. The experimental device enables attaching each slab either on the balance or on the translation stage and then removing the spacers while keeping unmodified the parallelism and the 7.7 mm gap. All the results presented in this paper are performed at the constant shear rate $\dot{\gamma}$ = 1.3·10$^{-3}$ s$^{-1}$. The high adhesive strength of gelatin avoids any wall sliding during the test [15]. Thanks to the transparent slab and image analysis, we measure the surface area of each sample, allowing for the stress to be plotted as a function of the strain. The stress-strain curve on Fig. 3 illustrates that the strain is small enough to remain in the linear elastic regime and that the shear modulus $G(\phi_p, d_p)$ can be simply deduced from the slope of the linear fit over a strain of 2.6 %. As an example, we can see on Fig. 3 how 5% of 20 µm particles strengthen the reference foam.

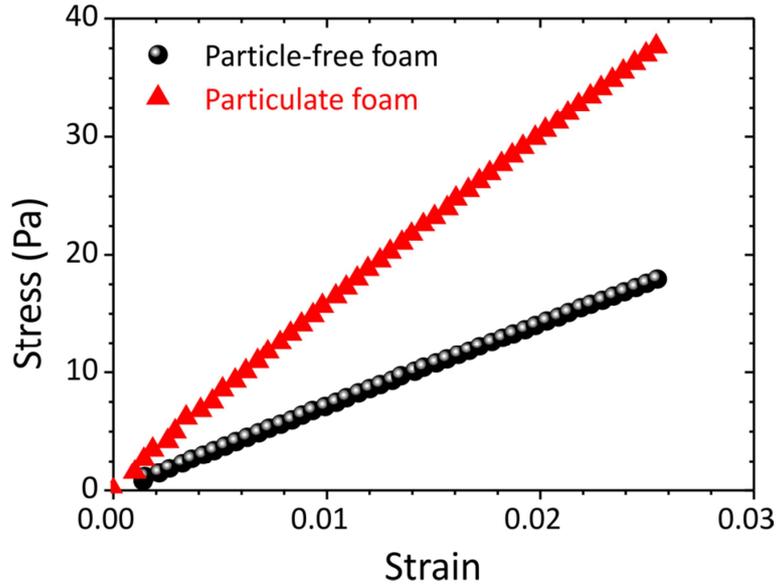

Fig. 3. Stress-strain measurements illustrating the strengthening of a foam by adding solid particles. From linear fits, the shear modulus is $G_0$ = 700Pa for the particle-free foam and $G$ = 1450Pa for a foam loaded with polystyrene particles ($d_p$ = 20 µm – $\varphi_p$ = 5 %).

### 3. Results and discussions

The influence of the particle loading will be characterized by normalizing the shear modulus of particle-laden foams by the value for the corresponding particle-free foam, i.e. $G^* = G(\phi, \varphi_p)/G(\phi, 0)$. Fig. 4 shows the evolution of the normalized shear modulus as a function of the particle volume fraction for several particle sizes, where two behaviors can be distinguished. The shear moduli of foams loaded with particles of diameter below 40 µm follow the same increasing function of $\phi_p$. Quantitatively, when particles are small enough compared to the bubble size, the shear modulus of particulate foam is roughly 3 times larger than the one of particle-free foam when only 6% of particles are added to the foam. In contrast, when the particle size increases, the stiffening can become negligible, as observed for the 250 µm particles. We can even notice a weak softening effect for the 500 µm particles. For that case, note that particle size becomes comparable to the bubble size.

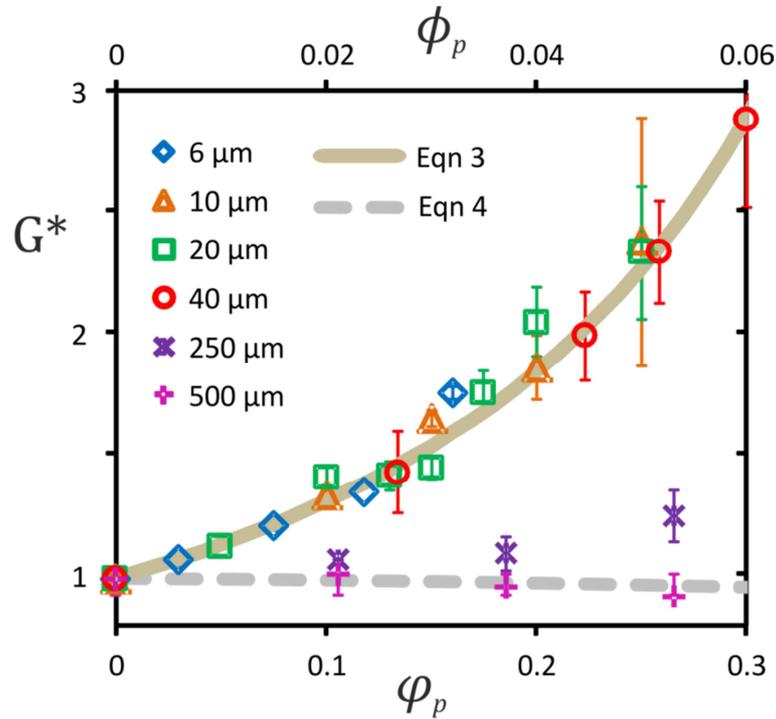

Fig. 4. The shear modulus of particle-laden foams normalized by the modulus of the corresponding particle-free foam as a function of the particle volume fraction in the interstitial volume $\varphi_p$ or the corresponding fraction in the whole sample volume $\phi_p$. Each symbol corresponds to different particle size as indicated in the legend. The solid line corresponds to Eq. (2) and the dashed one to Eq. (3). The errors bars represent the fluctuations obtained on four measurements.

To understand this stiffening-softening transition, it is necessary to consider the foam at the scale of the solid skeleton, or foam network. The size of interest is $d_c$, the diameter of the circle inscribed inside the minimal cross section of the struts, i.e. the constrictions (Fig. 5). In order to compare the particle size with $d_c$, we will use the so-called confinement parameter $\lambda$ [16]:

$$\lambda = \frac{d_p}{d_c} = \frac{1 + 0.57(1-\phi)^{0.27}}{0.27\sqrt{1-\phi} + 3.17(1-\phi)^{2.75}} \frac{d_p}{D_b} \quad (1)$$

For $\phi = 0.8$, Eq. (1) gives $\lambda \simeq 8.5\, d_p/D_b$. In [16] this geometrical parameter has been determined from both experiments involving the trapping/release of a single particle in aqueous foams and numerical simulations of foam structures. When $\lambda < 1$, particles can be included everywhere in the solid skeleton without deformation whereas when $\lambda \gtrsim 1$, particle deform the bubbles surface. One can also imagine that for $\lambda \gg 1$ the particles cannot be included anymore in the space defined by the foam network but instead the network is restructured around them. Thus, for a given particle volume fraction, the resulting particle configuration in the foam network depends crucially on $\lambda$, which is expected to have a significant influence on the mechanical behavior.

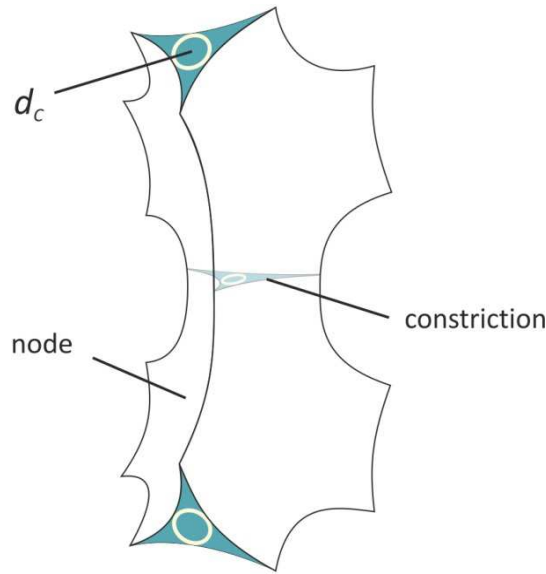

Fig. 5. Sketch of the foam network showing the nodes and the constrictions; $d_c$ is the diameter of passage through constrictions.

To quantify this influence, let's first consider the asymptotic case of small particles, i.e. $\lambda \ll 1$. The corresponding shear modulus will be called $G^*_{max}$. In one hand, the elastic modulus of solid foams normalized by the modulus of the solid skeleton is a decreasing function of the gas volume fraction $f(\phi)$ [17]. On the other hand, the modulus of an elastic matrix with rigid inclusions normalized by the modulus of the matrix is an increasing function of the particle fraction $g(\varphi_p)$ [18]. Thus, when the particles are small enough, the stiffness of the elastic skeleton is set by the particle volume fraction it contains. Thus, the reduced shear modulus of particulate foams can be written as a combination of those two functions:

$G^*_{max} = \frac{G(\phi,\varphi_p)}{G(0,\varphi_p)} \frac{G(0,\varphi_p)}{G(0,0)} \frac{G(0,0)}{G(\phi,0)} = f(\phi).g(\varphi_p).[f(\phi)]^{-1} = g(\varphi_p)$. In order to evaluate the function $g(\varphi_p)$ in the present foamy systems, we measured the reduced modulus of the gelatin matrix with several particle loadings (Fig. 6.a). We make use of the Krieger-Dougherty equation [19] to fit our data as suggested by previous studies on yield stress fluids loaded with particles [20]:

$$G^*_{max}(\varphi_p) = g(\varphi_p) \simeq \left(1 - \varphi_p/\varphi_p^*\right)^{-2.5\varphi_p^*} \quad (2)$$

where $\varphi_p^* \approx 0.6$. This phenomenological function can be used to estimate $G^*_{max}(\varphi_p)$, which is found to be in good agreement with experimental data for particles smaller than 40 µm, *i.e.* $\lambda \lesssim 0.84$, as shown in Fig. 4. This agreement can be considered as partly surprising because the underlying assumption of Eq. (2) is $\lambda \ll 1$. This indicates that a specific modeling should be developed to clarify the situations corresponding to $\lambda \lesssim 1$.

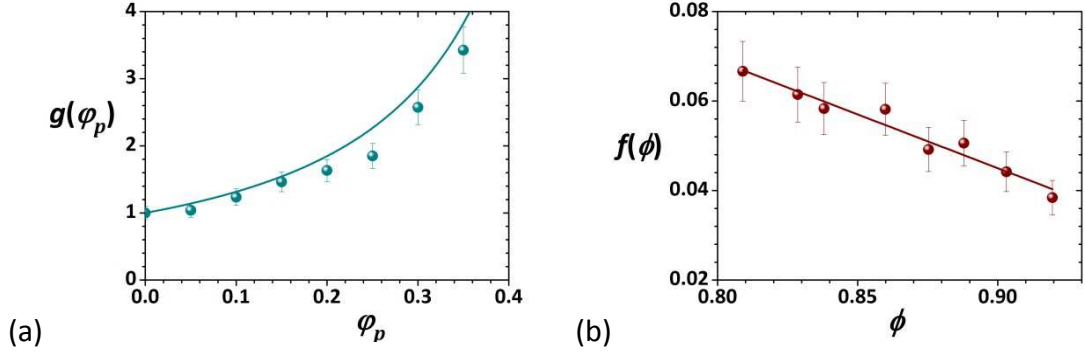

Fig. 6. (a) $g(\varphi_p) = \frac{G(0,\varphi_p)}{G(0,0)}$, the reduced shear modulus of a gelatin matrix with inclusions of polystyrene beads as a function of the particle volume fraction $\varphi_p$. (b) $f(\phi) = \frac{G(\phi,0)}{G(0,0)}$, the reduced shear modulus of a particle-free gelatin foam for different gas volume fraction $\phi$.

Let's now consider the case of $G^*_{min}$, which corresponds to the case of very large particles compared to the size of constrictions, i.e. $\lambda \gg 1$. That situation should be understood as particles large enough to be fully excluded from the space defined by the solid network, as illustrated in Fig. 7 for $\lambda \approx 10$. In such a case, the system is composed of particle-free foam that embeds large inclusions. As the particles no more contribute to the volume of the network, the gas fraction of the embedding foam, $\phi'$, is therefore increased with respect to the gas fraction in the whole system, $\phi$. The relation between $\phi'$ and $\phi$ is: $\phi' = \phi/(1 - \phi_p)$. In combining the effect of $\phi_p$ on $\phi'$, with the effect of gas fraction on the shear modulus of foams, one can deduce that the shear modulus of the embedding foam decreases as $\phi_p$ increases. Consequently, two opposing effects interact for particle-laden foams characterized by $\lambda \gg 1$: the particles stiffen the embedding foam whose shear modulus is decreased with respect to the corresponding unloaded foam. The global behavior of $G^*_{min}$ depends on the magnitude of those two effects. Similarly to what has been done for $G^*_{max}$, one can write

$\frac{G(\phi,\phi_p)}{G(\phi,0)} = \frac{G(\phi,\phi_p)}{G(0,\phi_p)} \frac{G(0,\phi_p)}{G(0,0)} \frac{G(0,0)}{G(\phi,0)}$, and in the limit case of $\lambda \gg 1$, $\frac{G(\phi,\phi_p)}{G(0,\phi_p)} = f(\phi')$.

Therefore, the corresponding global shear modulus is:

$$G^*_{min}(\phi_p) = g(\phi_p) \cdot \frac{f\left(\frac{\phi}{1-\phi_p}\right)}{f(\phi)} \quad (3).$$

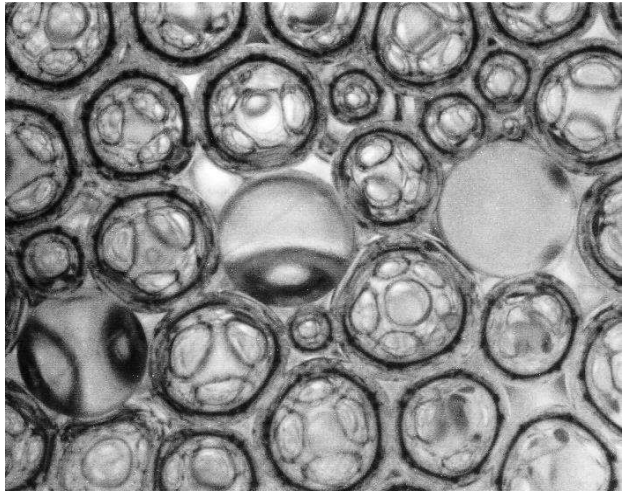

Fig. 7. Close-up on a gelatin foam loaded with 500 μm particles ($\lambda \approx 10$).

Here, the function $f$ has to be determined in order to evaluate $G^*_{min}(\phi_p)$, and we performed shear tests with particle-free gelatin foams, as shown in Fig. 6.b. Within the limited range of gas fraction investigated in the present study, we conveniently approximate the function $f(\phi)$ by $f(\phi) \simeq 0.26 - 0.24\phi$. With functions $f$ and $g$, one can estimate $G^*_{min}(\phi_p)$. The corresponding values are plotted in Fig. 3 against experimental data, showing good agreement. In particular, Eq. (3) predicts the global softening of the foamy material, in spite of the presence of solid inclusions.

Note that the two assumptions presented above have been considered earlier by [7], who have predicted the corresponding mechanical behavior using a self-consistent homogenization scheme. Although their particles samples were significantly polydisperse, the results obtained for "big" carbonate fillers [7] were captured by the assumption corresponding to $G^*_{min}$ in our paper, whereas the behavior induced by fillers of smaller size seem to be intermediate between $G^*_{min}$ and $G^*_{max}$. In contrast to those results, our data highlight a clear transition between the two asymptotic cases described above. This can be emphasized by setting $\varphi_p$ and plotting $G^*$ as a function of $\lambda$ (inset of Fig. 8). The transition observed for $\lambda \gtrsim 1$ can be interpreted in terms of morphological evolution associated to the progressive exclusion of the particles from the foam skeleton towards totally excluded particles for $\lambda \gg 1$. The modeling of the mechanical consequences of such a morphological evolution is challenging, and this aspect would certainly deserve a dedicated work. Here, we refer to a previous work [21] dealing with the drainage of foamy granular suspensions. At first sight, there is no direct comparison between foam drainage and the foam mechanics but, as we will show in the following, those two issues involve the same morphological transition. Indeed, in [21] the viscous drag (resistance) experienced by the draining liquid was measured to be maximal ($R_{max}$) when trapped particles were fully included in the foam network, i.e. $\lambda \approx 1$, but minimal ($R_{min}$) when the particles were fully excluded from the network, i.e. $\lambda \gg 1$ (see supplementary information). In introducing a function $\chi$ that measures the level of viscous drag between those two bounds, i.e. $\chi = (R(\lambda) - R_{min})/(R_{max} - R_{min})$, the corresponding drainage data were reasonably described by the phenomenological functional form:

$$\chi = 2.65\lambda^{-7/4} \quad (4).$$

In order to show that the transition observed for the shear modulus originates from the same morphological transition, we introduce a similar function: $\tilde{G} = [G^*(\lambda) - G^*_{min}]/[G^*_{max} - G^*_{min}]$, where $G^*_{max}$ and $G^*_{min}$ are respectively calculated with Eqs. (2) and (3). $\tilde{G}$ and $\chi$ are plotted in Fig. 8 and show that both the drainage of foamy granular suspensions and the shear modulus of particulate elastic foams undergo the same transition, which is controlled by $\lambda$ and which accounts for the evolution of the particle configuration in the foam. This highlights the generic nature of the reported transition, which concerns more physical properties of particle-laden foams than mechanics. Moreover, from a practical point of view, drainage and mechanics are intimately linked insofar as before being solid, particle-laden foams are liquid and they undergo the effects of drainage.

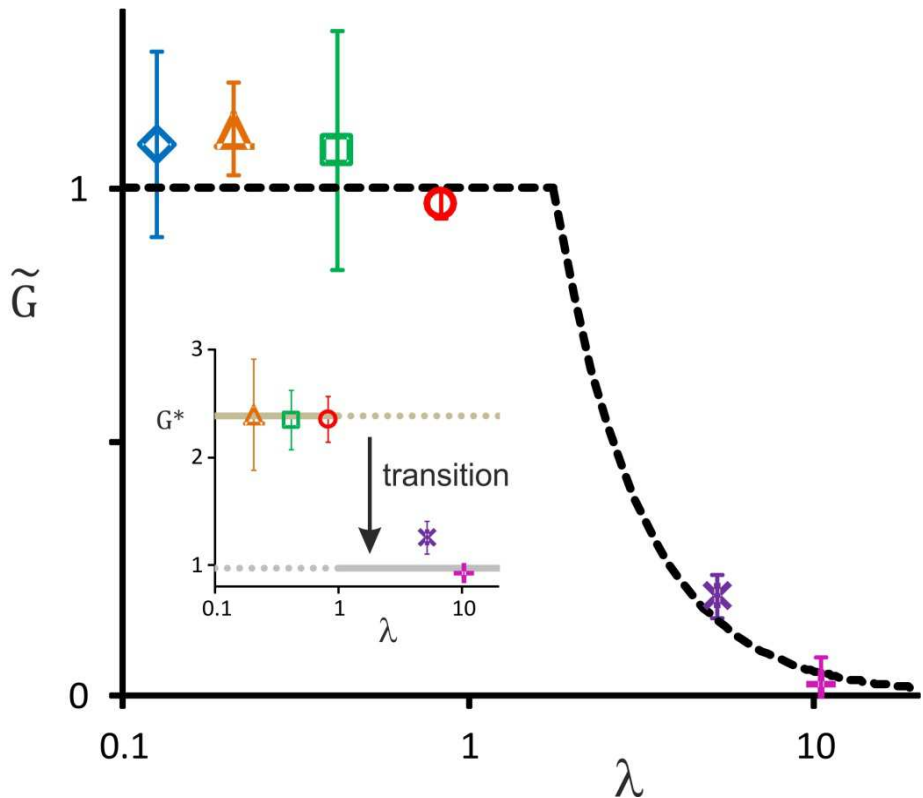

Fig. 8. The rescaled shear modulus of particulate gelatin foam for $\varphi_p$=26%. (same symbols than those on Fig 4) is compared to Eq. 4 obtained from data on the hydrodynamic resistance an aqueous foam loaded with particles for $\lambda > 1$. For particulate gelatin foam, $\lambda$ highlights a transition from a stiffening regime for small particles to a softening regime for large particles in the same $\lambda$-range than for the resistance of aqueous foam. Inset: Reduced shear modulus of particulate gelatin foam *vs* $\lambda$ for $\varphi_p$=26%.

## 4. Conclusion

We have addressed the issue of the strengthening of foamy materials by hard inclusions. The shear elastic modulus of particle-laden polymer foams has been measured as a function of the particle loading, in such a way that the ratio $\lambda$ of particle size to the foam network typical size was controlled. This geometric parameter was found to have a crucial influence on the mechanical behavior. For $\lambda < 1$, every particle loading leads to a strengthening effect whose magnitude depends only on the particle volume fraction. On the contrary, for $\lambda > 1$, the strengthening effect weakens abruptly as a function of $\lambda$, and a softening effect was even observed for $\lambda \gtrsim 10$. This transition in the mechanical behavior of foamy composite materials has been interpreted in terms of the evolution for the particles configuration in the foam network. This morphological evolution, the so-called "*particle exclusion transition*", has been recently highlighted in the framework of the drainage of foamy suspensions and it seems to be also relevant for the mechanics of such foamy systems.

**Acknowledgements**

We are grateful to G. Noiret and S. Audoly from Rousselot who kindly supplied the gelatin powder. We thank D. Hautemayou and C. Mézière for technical support. We gratefully acknowledge financial support from Agence Nationale de la Recherche (Grant no. ANR-13-RMNP-0003-01) and French Space Agency (convention CNES/70980).